\documentclass[aps,prl,twocolumn,showpacs,preprintnumbers,groupedaddress]{revtex4-1}

\usepackage{natbib}
\usepackage{graphicx}
\usepackage{dcolumn}
\usepackage{bm}

\begin{document}


\title{X-ray Raman Scattering from Water Near the Critical Point:\\ Comparison of an Isotherm and Isochore}


\author{D. Ishikawa$^{1,2}$}
\author{Y.Q. Cai$^{3}$}
\author{D.M. Shaw$^{4}$}
\author{J.S. Tse$^{4}$}
\author{N. Hiraoka$^{5}$}
\author{A.Q.R. Baron$^{1,2}$}
\affiliation{
$^{1}$Materials Dynamics Laboratory, RIKEN SPring-8 center, 1-1-1 Kouto, Sayo-cho, Sayo-gun, Hyogo, 679-5198, Japan\\
$^{2}$Research $\&$ Utilization Division, Japan Synchrotron Radiation Research Institute, 1-1-1 Kouto, sayo-cho, Sayo-gun, Hyogo, 679-5148, Japan\\
$^{3}$Photon Sciences, Brookhaven National Laboratory, PO Box 5000, Upton, NY 11973\\
$^{4}$Department of Physics and Engineering Physics, University of Saskatchewan, Canada \\
$^{5}$National Synchrotron Radiation Research Center, Hsinchu 30077, Taiwan
}


\date{\today}

\begin{abstract}
X-ray Raman spectra of liquid, sub- and super- critical water at the oxygen $K$-edge were measured, at densities $\rho$ = 1.02 - 0.16 gcm$^{-3}$.  
Measurements were made along both an isotherm and an isochore passing near the critical point.  
As density is reduced there is a general tendency of the spectra to increasingly resemble that of the vapor phase, with, first, a well separated low-energy peak, and, eventually, at densities below the critical density, 
peaks appearing at higher energies corresponding to molecular transitions.  
The critical point itself is distinguished by a local maximum in the contrast between some of the spectroscopic features. 
The results are compared to computed X-ray absorption spectra of supercritical water. 
\end{abstract}



\maketitle

\begin{figure}[!t]  
  \begin{center} 
    \begin{tabular}{cc} 
      \resizebox{43mm}{!}{\includegraphics{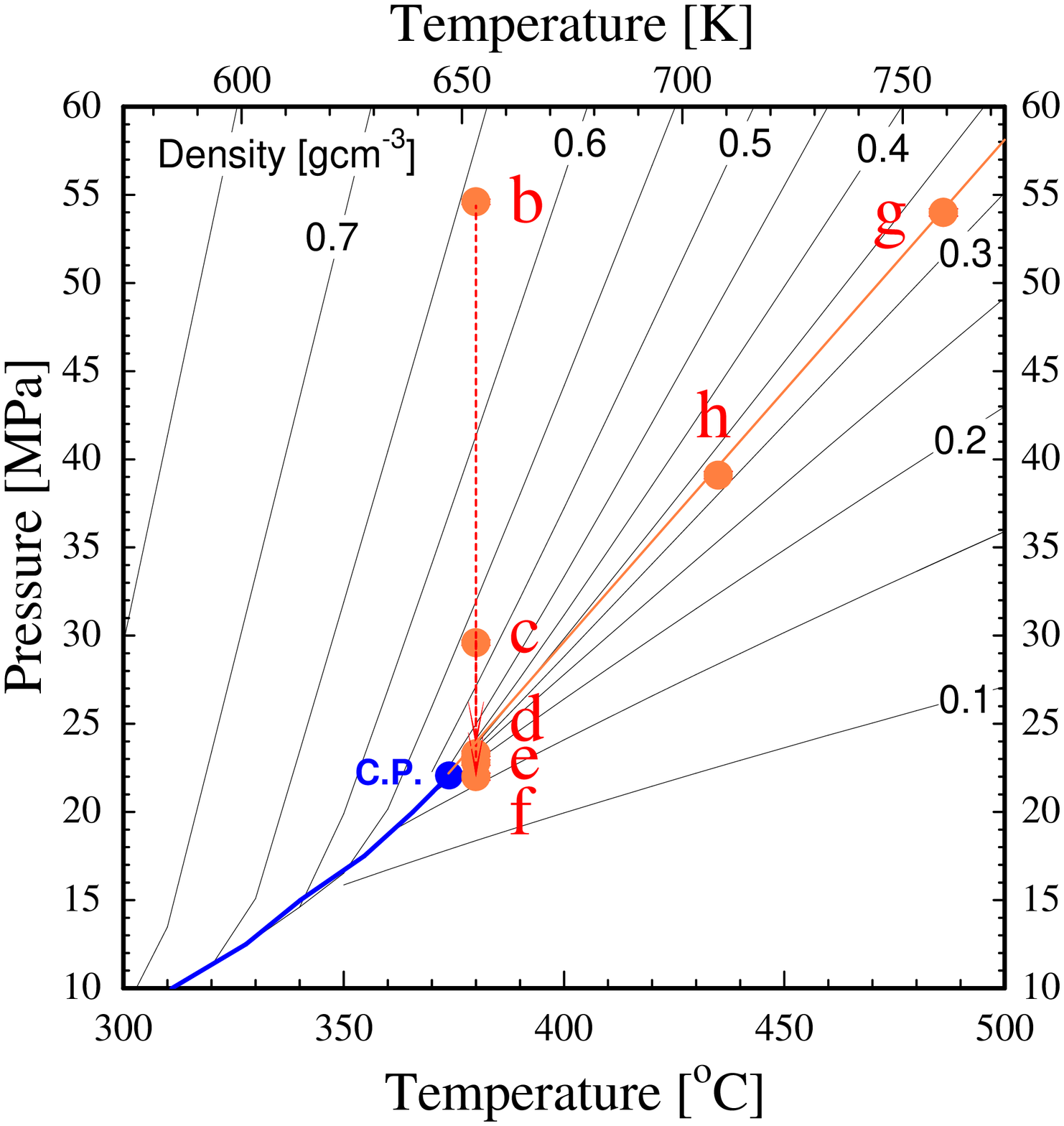}} &
      \resizebox{43mm}{!}{\includegraphics{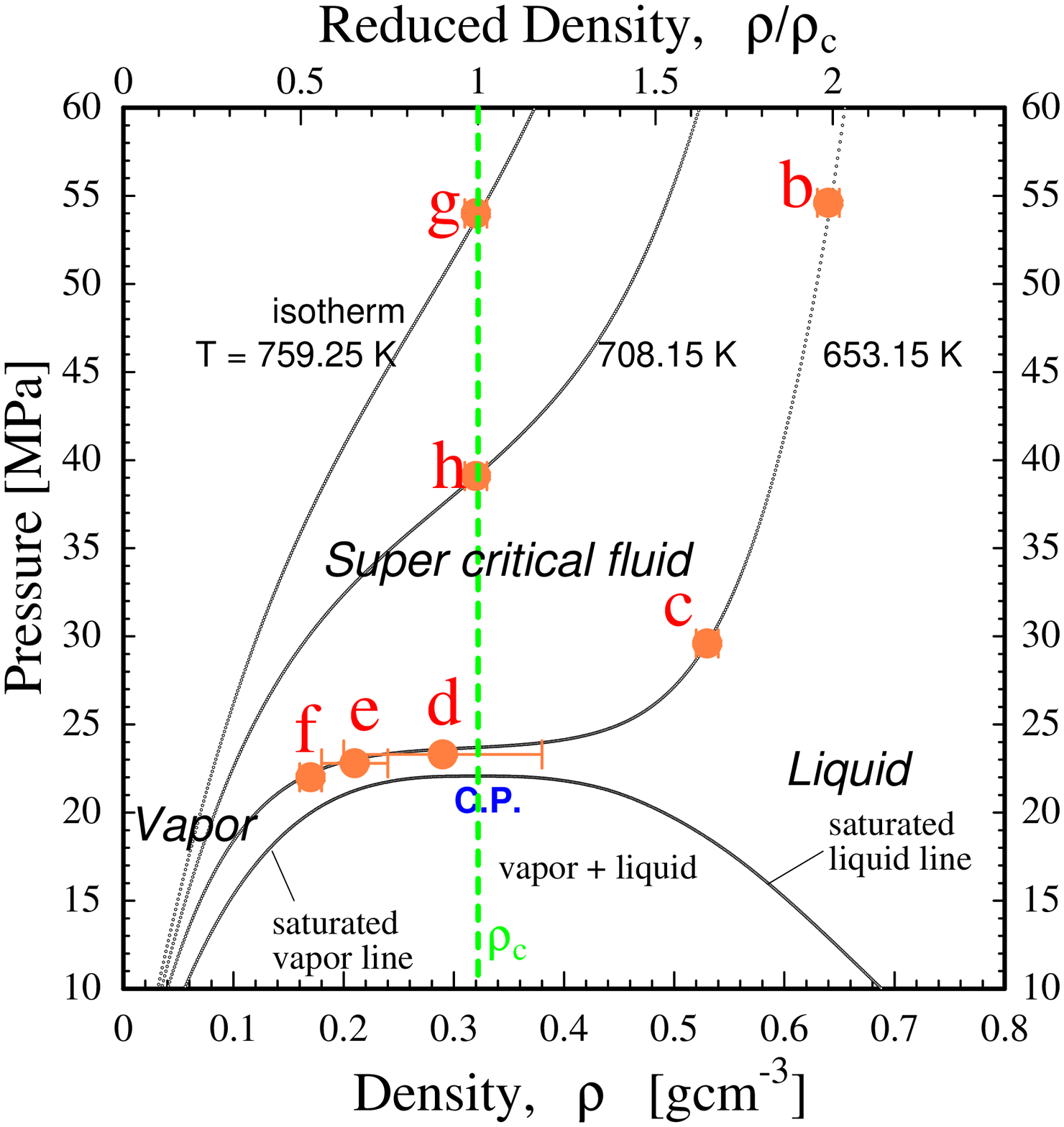}} \\
    \end{tabular}
\caption{\label{MP1}
Left: (T-P) phase diagram and isochores of H$_2$O \cite{wanger2002}. 
The solid line corresponds to the liquid-gas coexistence curve.  
Dashed lines indicate the expanded process and solid dots indicate the measured conditions. 
Right: ($\rho$-P) phase diagram and isotherm of H$_2$O. 
The dashed line indicates critical isochore.}
\label{MP} 
  \end{center}
\end{figure}

Heating a liquid under pressure allows one, eventually, to pass into the super-critical region where there is a continuous change in density between liquid and vapor phases.  
The neighborhood of the last point where there is a discontinuous change between the two phases, the critical point (see Fig.\ref{MP1}(left)), is interesting for a variety of reasons.  
Fundamentally, the critical point is where strong density fluctuations occur on essentially all length scales, leading to unique physical properties. Practically, the properties of liquid solvents in the supercritical region facilitate different types of reactions, often without the (sometimes environmentally harmful) catalysts that can be needed in other conditions \cite{franck87,bruno91,kiran94,arai2002}.

Water in the critical and supercritical region (critical point, T$_{\mathtt c}$ = 647.096 K, P$_{\mathtt c}$ = 22.064 MPa, $\rho_{\mathtt c}$ = 0.322 gcm$^{-3}$ \cite{wanger2002}) has attracted significant attention both for fundamental
and practical reasons.  On the fundamental side, there is great interest in the nature of the bonding in this region: how do the hydrogen bonds, which are responsible for many of the interesting features of liquid water in ambient conditions, survive into the super-critical region?  
There have been several studies focusing on this topic (e.g. experimentally \cite{postorino93,tromp94,bruni96,soper97,bellissent-funel97,botti98,tassaing98,ricci98,carey98,okada99,matsubayashi2001,wernet2005,sit2007} and theoretically \cite{mountain89,chummings91,fois94,mizan96,kalinichev97,jedlovszky98,boero2000,marti2000}).   
However, there remains significant debate and uncertainty, in part because experiments on water in this region (where, for example, its reactivity is sufficiently high to etch typical stainless steel) are difficult, and in part because the interpretation of spectroscopic experiments is severely complicated by the density fluctuations.

In the present paper we apply X-ray Raman Scattering (XRS) to investigate the behavior of water in the critical region and compare our results to ab-initio calculations.    
Soft X-ray absorption spectroscopy (XAS) \cite{stohr92,schulke92} and XRS \cite{mizuno67} have revealed the details of hydrogen bonding (HB) environments in water by assigning the observed spectral features near the oxygen $K$-edge \cite{bowron2000,bergmann2002,cavalleri2002,wernet2004,cavalleri2004,cai2005,wernet2005}. 
(Non-resonant) XRS is alternative to XAS. 
Under the condition $qr_0\ll 1$ ($r_0$: radius of core state), the dipole contribution is dominant and has the same matrix element as absorption. 
In this case, the spectral shape of XRS is proportional to XAS cross section. 
The higher energy of XRS experiments (6 - 12 keV), as compared to XAS work ($\sim$ 540 eV), means that it is much easier to penetrate into complex sample environments, as is very important for investigating supercritical water (SCW).  

The experiments were performed on the Taiwan beamline BL12XU \cite{cai2004} at SPring-8 in Japan.  
Incident radiation was monochromatized by a Si(400) reflection  ($-,+,+,-$) 4-bounce high-resolution monochromator and focused onto the sample to a spot size of 80 (V) $\times$ 120 (H) $\mu$m$^2$.
The scattered radiation is analyzed using a 2-m radius spherically bent Si(555) analyzer crystal in a Rowland circle geometry.  
The analyzer angle of was fixed at 88.5$^{\circ}$, corresponding to an analyzer energy of 9888.8 eV.
The incident photon energy was scanned from 10421 to 10437 eV.  
The total energy resolution was 260 meV (FWHM of the quasi-elastic line of the sample) measured at the analyzer energy. 
The energy scale was calibrated using scans of the Tantalum $L_{3}$ (9881.1 eV) and Rhenium $L_{3}$ (10535.3 eV) absorption edges.

The sample cell was custom designed for inelastic X-ray scattering experiments, with an inner chamber made of Hastelloy-X alloy with diamond windows to avoid reaction with the SCW.  For this work, we chose a 3 mm sample length, corresponding to the $1/e$ absorption length for water of density 0.6 gcm$^{-3}$.  The incident and outgoing angular acceptances were 22$^{\circ}$.  This cell was then placed inside of a vacuum chamber (to aid thermal control and reduce air-scatter) with polyimide windows.

The sample was ultra-pure water.  Its thermodynamic conditions were maintained by heating the inner sample cell and providing pressure using an external hand press.  As the sample density is extremely sensitive to temperature and pressure especially near the critical point, the pressure was continuously monitored throughout the experiment and the temperature was controlled using an Inconel-covered thermocouple in direct contact with the water in the cell (just to one side of the X-ray path).   Both pressure and temperature measurement systems were carefully calibrated, with the temperature gauge expected to be accurate to 0.4 \% and the pressure to $\pm$ 0.1 MPa and 0.1 \%.  During scans, measured temperatures were maintained to $\pm$ 0.1 K and pressures to $\pm$ 0.05 MPa.

Measurements were made at an 18$^{\circ}$ scattering angle, or $q \simeq$ 15.7 nm$^{-1}$.  This satisfies the dipole scattering approximation with $qr_1 \sim 0.1 \ll 1$ ($r_1 = a_0/Z$  is the radial extent of the oxygen 1s wave function, where, $a_0$ is Bohr radius and $Z$ is effective nuclear charge for the orbital).  
  The acceptance of the analyzer was 2.5 nm$^{-1}$.   Scans over the full energy range, 532 - 548 eV energy transfer were typically 90 minutes long (including a check of the elastic peak) and depending on the conditions, e.g. sample density, typically 10 to 20 scans were measured at each set of thermodynamic conditions, with, sometimes, a shorter range chosen to efficiently use beam time.  For each scan, the signal was normalized by the incident intensity and backgrounds were subtracted assuming an exponential energy dependence, i.e. primarily, due to the Compton tail of the diamond windows.   The spectra were then normalized to unity integral over 532 to 548 eV energy transfer, with a smooth continuation used if the measured spectra did not cover the full range.

\begin{figure}
\includegraphics[width=8.cm]{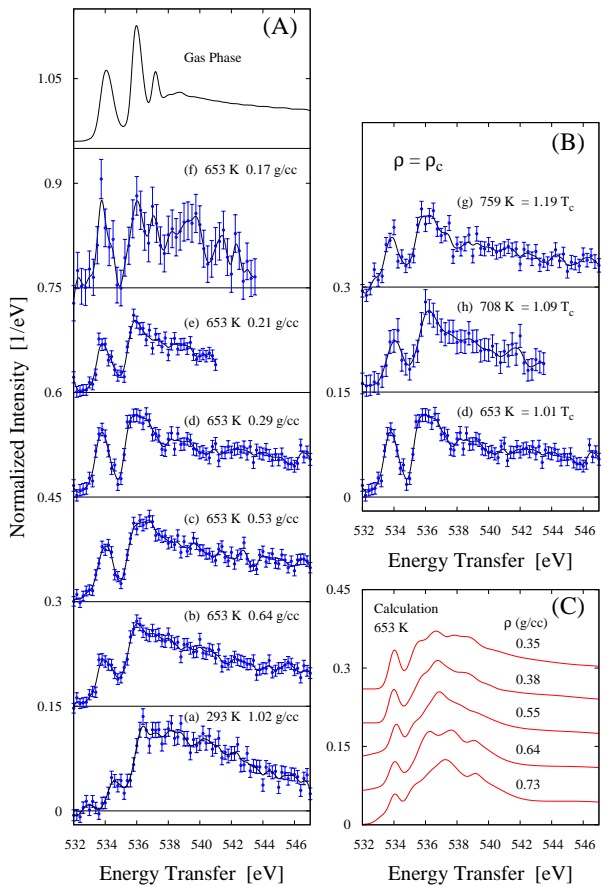}
\caption{\label{XRS_SCW_data} 
(a)-(h): Area normalized XRS spectra in expanded water; near $K$-edge spectra of the oxygen 
obtained with a total energy resolution of 260 meV (FWHM) and 0.2 eV step.
Open circles with statistical error bars are measured data and solid lines are smoothed curve.
The thermodynamic conditions are indicated in Table \ref{tab1}. 
(Gas Phase): Dilute gas phase spectrum (fluorescence-yield soft-XAS from Ref.\cite{schirmer93}) convoluted by resolution function of this work.
(A) along with isotherm T = 1.01 T$_{\mathtt c}$ 
(B) along with critical isochore $\rho=\rho_{\mathtt c}$.
(C) Computed XAS spectra of SCW for several densities at T = 653 K. 
See text and supplemental material.
}
\end{figure}

The measured conditions are indicated on the phase diagram in Fig.\ref{MP}, with the precise conditions given in Table \ref{tab1}. Fig.\ref{XRS_SCW_data}(A) and (B) present the measured spectra. 
 The XRS/XAS spectra can be divided into three regions: (I) pre-edge (533-535 eV), (II) main-edge (535-538 eV), and (III) post-edge ($>$ 538 eV). 
Fig.\ref{XRS_SCW_data}(A) shows the isothermal response at the critical temperature and Fig.\ref{XRS_SCW_data}(B) the response along isochore at the critical density, both passing close to the critical point (spectrum (d)).  
The clearest trend is visible along the isothermal line, showing a gradual progression from the relatively blurred out spectra at high density (b), similar to bulk water, to something closer to the gas-phase spectra at low density (f).  
There is also a shift, by about 0.5 eV, to lower energy of the pre-edge when the temperature is increased from 293 K to 653 K (spectrum (a) to (b)).  
The lowest energy peak corresponds, in the gas phase, to the 1s $\to$ 4a$_1$ transition of the H$_2$O molecule, which has an excited state wave function spread out over both hydrogen atoms and separate portion about the oxygen atom.  
This peak becomes more distinct with decreasing density or pressure.  
Meanwhile there are also changes in the high-energy part of the spectrum, but these are hard to quantify, especially in the absence of a rather good model.  
The trends in the isochore, varying from 1.01 to 1.19 T$_\mathtt{c}$ , are not so strong. 

\begin{table}
\caption{\label{tab1}
\footnotesize
Thermodynamic conditions and uncertainties at which the XRS spectra of expanded water were measured. 
The spectra are recorded along one near-critical isotherm T $\sim$ 1.01 T$_{\mathtt c}$ (b, c, d, e, f) and along one critical isochore $\rho$ = $\rho_{\mathtt c}$ (g,h,d). 
Labels indicate in Fig. \ref{MP}
}
\begin{ruledtabular}
\begin{tabular}{c|cc|c|cc|l}
\footnotesize
 & T (K) & P (MPa) & $\rho^{(1)}$(gcm$^{-3})$ & $\rho_+$ & $\rho_-$ & $\rho^{(2)}$(gcm$^{-3})$ \\
\hline
a & 293.15(1.6) & 54.6(0.15) & 1.02(0.00)& 1.02 & 1.02 & 1.02(0.00)  \\
b & 653.15(1.6) & 54.6(0.15) & 0.64(0.01) & 0.64 & 0.65 & 0.64(0.00)  \\
c & 653.15(1.6) & 29.6(0.15) & 0.53(0.01) & 0.52 & 0.54 &  0.53(0.01) \\
d & 653.15(1.6) & 23.3(0.15) & 0.29(0.09) & 0.21 & 0.34 &  0.24($^{+0.14}_{-0.04}$) \\
e & 653.15(1.6) & 22.8(0.15) & 0.21(0.03) & 0.18 & 0.25 &  0.20($^{+0.06}_{-0.02}$) \\
f & 653.15(1.6) & 22.0(0.20) & 0.17(0.01) & 0.16 & 0.17 &  0.16($^{+0.02}_{-0.01}$) \\
g & 759.25(2.0) & 54.0(0.20)  & 0.32(0.01) & 0.32 & 0.33 &  0.32(0.01)  \\
h & 708.15(1.8) & 39.1(0.20)  & 0.32(0.01) & 0.31 & 0.33 &  0.32(0.01)  \\  
\end{tabular}
\end{ruledtabular}
\footnotetext{
T: nominal  temperature, P: nominal  pressure, $\rho^{(1)}$: center of the error value at T, $\rho_+$: upper limit of the error at T+$\Delta$T, $\rho_-$: lower limit of the error at T-$\Delta$T, $\rho^{(2)}$: nominal value
}
\end{table} 
\begin{figure}
\includegraphics[width=8cm]{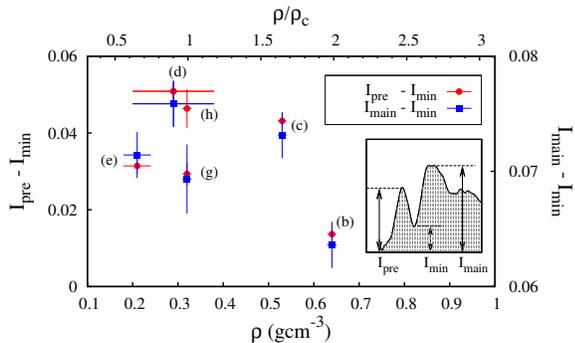}
\caption{\label{enhance} 
Density dependent contrast of pre-edge with maximum near the critical point. 
}
\end{figure} 

The spectrum at the critical point, (d), is distinguished from the others by having a relatively strong contrast between the low-energy (presumably 1s $\to$ 4a$_1$) peak and the high-energy part of the spectrum.   This is brought out in more detail in Fig.\ref{enhance}, where the difference between the maximum and minimum intensities is plotted.   The contrast in spectrum (d) is stronger than that in spectrum (c) and (e) (at T$_{\mathtt c}$  but, respectively, at higher and lower pressure) and (h) on the high-temperature side of the critical isochore.  This contrast is due to an increase in height of the low-energy peak relative to the high-energy edge.  
If one considers the response in the critical region to be the sum over the spectral response of clusters of various sizes, improved contrast in the conditions where the largest distribution of cluster sizes is expected is surprising.  
Thus these measurements might indicate that we are seeing effects from electronic motions on time scales comparable to the XRS.

Our data is noticeably different than a recent publication \cite{wernet2005} of data in nominally similar thermodynamic conditions.  
This difference might be the result of a large non-dipole contribution in Ref.\cite{wernet2005} (they had $qr_1\sim$ 0.45) or other experimental differences (they had somewhat worse, $\sim$1 eV energy resolution and oxygen-containing sapphire windows).  
If we apply their scaling argument, the size of the peak at 534 eV suggests about 73(10) \% of the molecules are nearly gas-like in spectrum (c) (the point that was closest to the conditions stated in Ref.\cite{wernet2005}) and about 85(10) \% in spectrum (d), a much-larger gas-like percentage than the 35(20) \% of Ref.\cite{wernet2005}.  
However, such a scaling argument probably over-simplifies the problem and detailed simulations are really required to interpret the results.

Simulations were performed using Car-Parrinello molecular dynamics (CPMD) \cite{cpmd} in the microcanonical ensemble with density functional theory.  
Initial structures were obtained from the DL\_POLY \cite{smith96} simulations for densities 0.73, 0.64, 0.55, 0.38 and 0.35 gcm$^{-3}$ at 653 K using the TIP2P water model. 
Exchange-correlation function \cite{perdew96} and norm-conserving pseudopotentials \cite{troullier91} were used.  
Oxygen $K$-edge XAS spectra were computed using the transition state potential approach of Slater \cite{slater74} for the geometries obtained in the molecular dynamics runs.  
Details of the computation are presented in the supplementary materials.

The calculated XAS along isotherm T = 653 K are shown in Fig.\ref{XRS_SCW_data}(C).  
A comparison of the calculated XAS of SCW and vapor-phase from experiment shows the presence of the pre-edge peak in all of the spectra at a similar excitation energy (534 eV).  
The calculated pre-edge peak height increases with the decrease in density consistent with the trend of experimental results. 
The main-edge region narrows with the decrease in density in both experiment and calculation, though this is not clearly consistent especially near the critical density.

\begin{figure}[!t]
\includegraphics[width=8cm]{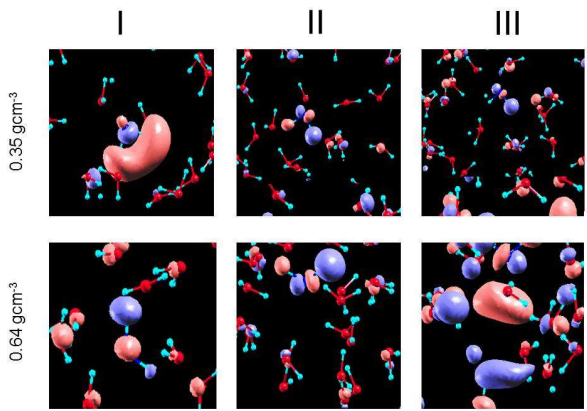}
\caption{\label{waveplot} 
Molecular orbital plots for the final states with respect to the three excitation regions (I: pre-edge, 533-535 eV; II: main-edge, 536-538 eV; III: post-edge, $>$ 538 eV).  
The two thermodynamic conditions, $\rho$ = 0.64 gcm$^{-3}$ (= 2.0 $\rho_c$) and $\rho$ = 0.35 gcm$^{-3}$ (= 1.09 $\rho_c$), are selected.  
H$_2$O molecules are shown as v-shaped sticks with ball.  
The atoms color code is red for O and light blue for H. 
The molecular orbitals are plotted in positive (pink) and negative (purple) valued wavefunctions using isosurface values of (region I) 0.05 $e$\AA$^{-3}$; (region II) 0.02 $e$\AA$^{-3}$; (region III) 0.015 $e$\AA$^{-3}$.            
}
\end{figure}

Examination of the final state wavefunction (Fig.\ref{waveplot}) indicates that the origin of pre-edge peak (I) is excitations to final states that are a combination antibonding OH and Rydberg orbitals.  
Examination of the orbital plots shows that these states are confined to the water molecule with the tail of the wavefunction extending only to nearest neighbor molecules.  
Thus, changes in the HB at the excited water molecule will have little effect on the shape and energies of these highly localized orbitals.  
This explains why the pre-edge peak position and profile are largely unaffected by the surrounding environment, showing no significant modification with a change in the density of SCW.   
Examination of orbital plots from the calculated XAS indicate that transition in the main-edge region (II) are to final states with antibonding OH orbitals and a larger mixture of Rydberg states than found in the lower-lying excitations that are responsible for the pre-edge peak (I).  
At lower densities, where less interactions with surrounding water molecules are expected, there is a lower mixing in the diffuse Rydberg states, resulting in an increase of the energies and a narrowing of the distribution of these states.  
This is the reason for the narrower main-edge peak in the 0.35 gcm$^{-3}$ spectrum compared to 0.64 gcm$^{-3}$. 
An examination of the post-edge region (III) orbital plots for SCW indicates that excitations in this region are to diffuse, continuum orbitals.  
A decrease in density shows some localization of the wavefunctions of these states between the molecules.  
Both measurements and calculation show sensitive changes to this region promising additional information if calculations are improved.  

In conclusion, measurements of the oxygen $K$-edge x-ray Raman scattering show marked changes as thermodynamic conditions are tuned in the neighborhood of the critical point.  
There is, broadly speaking, reasonable qualitative agreement with XAS calculations that allows us to interpret these results in terms of orbital structures.  
However, the contrast in the measured data is better than in the calculation, especially near the exact critical conditions, which hints that we may be seeing an effect of fast electronic dynamics in the XRS spectra.  
The complex high-energy structure of the spectra and calculations also suggest significantly more information might be obtained if the calculations can be modified to bring them into closer agreement with the data.

\begin{acknowledgments}
We would like to thank Dr. Yasuo Ohishi for loan of a diamond windows.  
The experiments were carried out at the SPring-8 under approvals with JASRI (No.2006A4255, 2007A4252) and NSRRC, Taiwan (No.2005-3-101-2, 2005-3-101-5).  
DI is supported by Grant-in-Aid for Young Scientists (B), Japan Society for the Promotion of Science (JSPS No.17740203).  
YQC is supported by the U.S. Department of Energy, Office of Basic Energy Science, under the Contract No. DE-AC02-98CH10886.
\end{acknowledgments}

\bibliography{SCW_XRS}

\end{document}